\documentclass[aps,prb,showpacs,groupedaddress,twocolumn]{revtex4}

\usepackage{graphicx}
\usepackage{graphics}

\begin{document}

\title{Spin Nernst effect and Nernst effect in two-dimensional electron systems}

\author {Shu-guang Cheng$^1$, Yanxia Xing$^1$, Qing-feng Sun$^{1,\ast}$, and X.
C. Xie$^{2,1}$}

\address{
$^1$Institute of Physics, Chinese Academy of Sciences, Beijing
100190, China\\
$^2$Department of Physics, Oklahoma State University, Stillwater,
Oklahoma 74078}

\begin{abstract}
We study the Nernst effect and the spin Nernst effect, that a
longitudinal thermal gradient induces a transverse voltage and a
spin current. A mesoscopic four-terminal cross-bar device having the
Rashba spin-orbit interaction (SOI) under a perpendicular magnetic
field is considered. For zero SOI, the Nernst coefficient peaks when
the Fermi level crosses the Landau Levels. In the presence of the
SOI, the Nernst peaks split, and the spin Nernst effect appears and
exhibits a series of oscillatory structures. The larger SOI is or
the weaker magnetic field is, the more pronounced the spin Nernst
effect is. The results also show that the Nernst and spin Nernst
coefficients are sensitive to the detailed characteristics of the
sample and the contacts. In addition, the Nernst effect is found to
survive in strong disorder than the spin Nernst effect does.
\end{abstract}

\pacs{72.15.Jf, 72.25.-b, 73.23.-b, 73.43.-f} \maketitle

The Hall-like effect, namely, a longitudinal force induces a
transverse current, has been a fascinating topic since the early
days of the condensed matter physics. The integer and the
fractional quantized Hall effects, two celebrating examples, have
been extensively investigated for the last three decades, but
remain to be active research fields till now. Recently, another
Hall-like effect, spin Hall effect, in which a longitudinal
voltage bias induces a transverse spin current due to the
scattering by magnetic impurities or due to the existence of a
spin-orbit interaction (SOI), has generated a great deal of
interest.\cite{ref1,ref2} Apart from a large number of theoretical
studies, several experimental investigations also made important
contributions to the field. Up to now, SOI has indeed been found
to be substantial in some semiconductors despite it is a
relativistic effect, and its strength can be tuned by the gate
voltage in the experiment.\cite{ref3,ref4} In particular, the spin
Hall effect has been detected experimentally by observing the
transverse opposite-spin accumulations near the two edges of the
sample.\cite{ref5,ref6}

The Nernst effect, a thermoelectric property, in which a
longitudinal thermal gradient induces a transverse current (or a
bias $\Delta V$ with open boundary) while under a perpendicular
magnetic field, is also a Hall-like effect. The thermoelectric
coefficients (including the Seebeck coefficient and the Nernst
coefficient) of electronic systems are known to be more sensitive to
the details of the density of states than the
conductance,\cite{ref7,ref8,ref9} and these detailed information of
the density of states is importance for the design of the electronic
devices. But the thermoelectric measurement is usually more
difficult to carry out than the conventional transport measurements,
particularly for low-dimensional systems or nano-devices.
Fortunately, because of the development of the micro-fabrication
technology and the low-temperature measurement technology in the
last two decades, the thermoelectric measurement in low-dimensional
samples has been feasible now.\cite{ref10,ref11} Recently, the
thermopower of the quantum dot was measured, and the results in the
Kondo regime show a clear deviation from the semiclassical Mott
relation.\cite{ref10,ref12} The Nernst effect in bismuth has also
been detected and the Nernst coefficient peaks at positions when
Fermi level crosses over the Landau levels (LLs).\cite{ref13}
Meanwhile, limited theoretical studies of the Nernst effect have
also appeared.\cite{ref14}

In this paper, we study the Nernst effect and spin Nernst effect in
a two-dimensional electron gas with a SOI and under a perpendicular
magnetic field $B$. For the first time, the spin Nernst effect, a
novel Hall-like effect, is investigated.  The spin Nernest effect
implies that a longitudinal thermal gradient $\Delta \mathcal{T}$
induces a transverse spin current. The spin Nernst coefficient
should be more sensitive to the details of the spin density of
states of the system than the spin Hall conductance, similar as
their electronic counterparts.\cite{ref7,ref8,ref9} We consider the
system as shown in Fig.1a, consisting of a square center region
connected to four ideal semi-infinite leads. A longitudinal thermal
gradient $\Delta \mathcal{T}$ is added between the leads 1 and 3.
This thermal gradient induces a transverse Hall voltage $V_H$ with
the open boundary condition under a perpendicular magnetic field
$B$; a transverse spin current $J_{sH}$ in the closed boundary
condition with a SOI. By using a tight-binding model and the
Landauar-Buttiker (LB) formula with the aid of the Green's
functions, the Nernst coefficient $N_e$ ($N_e\equiv V_H/\Delta
\mathcal{T}$) and spin Nernst coefficient $N_s$ ($N_s \equiv
J_{sH}/\Delta \mathcal{T}$) are calculated. Without a SOI, the
Nernst coefficient $N_e$ peaks when the Fermi level $E_F$ crosses
the LLs, and spin Nernst coefficient $N_s$ is absent, consistent
with the recent experimental findings.\cite{ref13} In the presence
of a SOI, each LL splits into two, consequently, each Nernst peak
splits into two peaks. Meanwhile, the spin Nernst effect emerges and
its coefficient $N_s$ exhibits a series of oscillatory structures.
The oscillation is enhanced with increasing SOI but is damped by a
large $B$. In addition, the Nernst effect is found to survive in
strong disorder than the spin Nernst effect does.

\begin{figure}
\includegraphics[bb=40 412 576 652, scale=0.50, clip=]{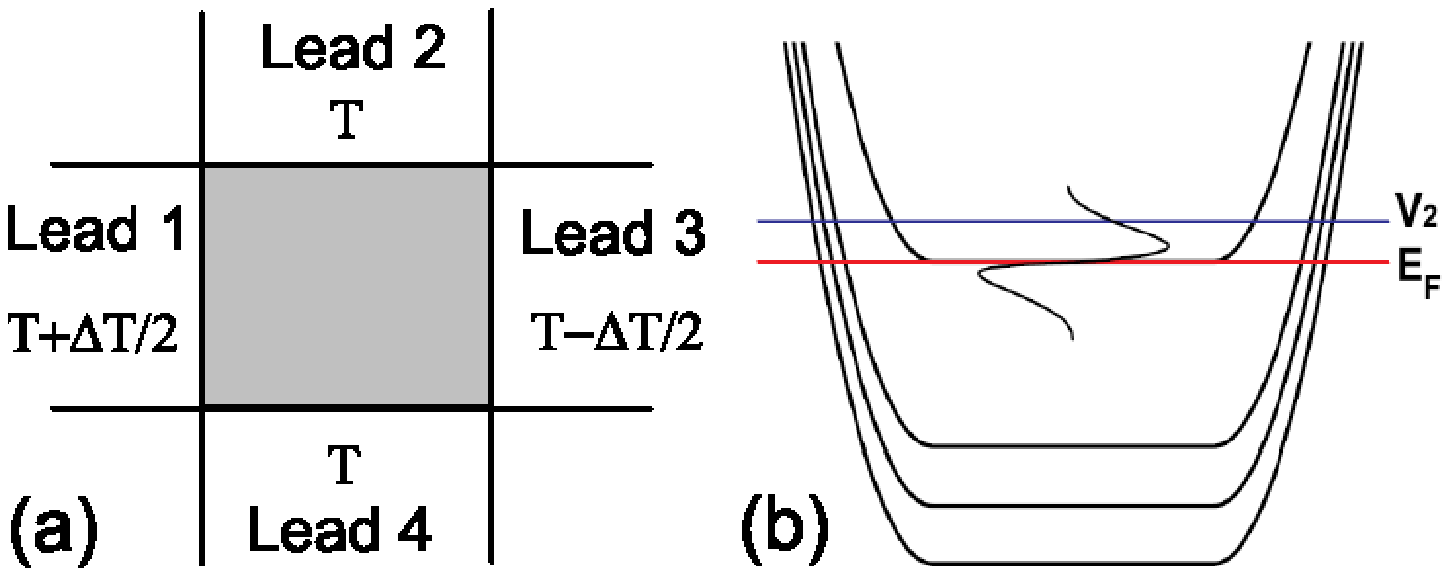}\\
\caption{(color online) (a) Schematic diagram for the four-terminal
cross-bar sample with a thermal gradient $\Delta \mathcal{T}$
applied between the longitudinal lead-$1$ and lead-$3$. (b)
Schematic view of the LLs of the center region, the Fermi energy
$E_F$, and the bias $V_2$. The oscillatory line across $E_F$ is the
difference of $f_1(E)-f_2(E)$. }
\end{figure}

In the tight-binding representation, the cross-bar sample is
described by the Hamiltonian:\cite{ref15},
\begin{eqnarray}
H&=&-t\sum_{{\bf i}\sigma}[c_{{\bf i}+\delta
x,\sigma}^{\dag}c_{{\bf i}\sigma}e^{-im\theta}+ c_{{\bf i}+\delta
y,\sigma}^{\dag}c_{{\bf i}\sigma} + H.c.] \nonumber\\
&+ &\sum_{{\bf i}\sigma}\varepsilon_{{\bf i}}c_{{\bf
i}\sigma}^{\dag}c_{{\bf i}\sigma} -V_R \sum_{{\bf
i}\sigma\sigma'}[c_{{\bf i}+\delta
y,\sigma}^{\dag}(i\sigma_x)_{\sigma\sigma'} c_{{\bf i}\sigma'}
\nonumber \\
&-&c_{{\bf i}+\delta x,\sigma}^{\dag}(i\sigma_y)_{\sigma\sigma'}
c_{{\bf i}\sigma'}e^{-im\theta}+H.c.]
\end{eqnarray}
where $c_{{\bf i}\sigma}^{\dag}$($c_{{\bf i}\sigma}$) is the
creation (annihilation) operator of electrons in the site ${\bf
i}=(n,m)$ with spin $\sigma$. $t=\hbar^2/(2m^*a^2)$ is the hopping
matrix element with the lattice constant $a$, and $\delta x$ and
$\delta y$ are the unit vectors along the x and y directions.
$\varepsilon_{\bf i}$ is the on-site energy, which is set to $0$
everywhere for the clean system. While in a disorder system,
$\varepsilon_{\bf i}$ in the center region is set by a uniform
random distribution [-W/2,W/2]. The last term in Eq.(1) represents
the Rashba SOI with $V_R$ being its strength. In order to avoid
confusion in calculating the spin current, $V_R$ is set to zero in
the lead-2 and lead-4. The extra phase $\theta ={ea^2}B/h$ is from
the perpendicular magnetic field $B$. Here the Zeeman effect and
electron-electron interaction are neglected.\cite{note1} The
Zeeman split could be small in some of the two-dimensional
electron systems. The electron-electron interaction is weak in
systems with high carrier density.

The particle current $J_{p\sigma}$ in the transverse lead-$p$ with
spin $\sigma=\uparrow,\downarrow$ can be obtained by the LB
formula:\cite{ref15}
\begin{equation}
J_{p\sigma}=\frac{1}{\hbar}\sum_{q\neq p} \int
dE~T_{p\sigma,q}(E)[f_p(E)-f_q(E)] \label{Landau2}
\end{equation}
where $T_{p\sigma,q}(E)$ is the transmission coefficient from the
lead-$q$ to the lead-$p$ with spin $\sigma$ and $E$ is the energy
of the incident electron. The transmission coefficient can be
calculated from: $T_{p\sigma,q}(E)=Tr[{\bf \Gamma}_{p\sigma}{\bf
G}^r{\bf \Gamma}_{q}{\bf G}^a]$, where the line-width function
${\bf \Gamma}_{p\sigma}(E)=i({\bf \Sigma}_{p\sigma}^r-{\bf
\Sigma}_{p\sigma}^{r\dagger})$, ${\bf \Gamma}_{q} ={\bf
\Gamma}_{q\uparrow}+{\bf \Gamma}_{q\downarrow}$, and ${\bf
\Sigma}_{p\sigma}^r$ is the retarded self-energy due to coupling
to the lead-$p$ with spin $\sigma$. The Green's function ${\bf
G}^r(E)=[{\bf G}^a(E)]^{\dagger}=\{E{\bf I}-{\bf
H}_0-\sum_{p\sigma}{\bf \Sigma}^r_{p\sigma}\}^{-1}$ and ${\bf
H}_0$ is the Hamiltonian of the central region. $f_p(E)$ in Eq.(2)
is the electronic Fermi distribution function of the lead-$p$, and
$f_p(E)=1/\{ {\rm exp}[(E-E_F-V_p)/k_B \mathcal{T}_p]+1\}$ with
the bias $V_p$ and temperature $\mathcal{T}_p$. After getting the
particle current $J_{p\sigma}$, the (charge) current is
$J_{pe}=e(J_{p\uparrow}+J_{p\downarrow})$ and the spin current is
$J_{ps}=(\hbar/2)(J_{p\uparrow}-J_{p\downarrow})$.

\begin{figure}
\includegraphics[bb=39 7 542 373, scale=0.50, clip=]{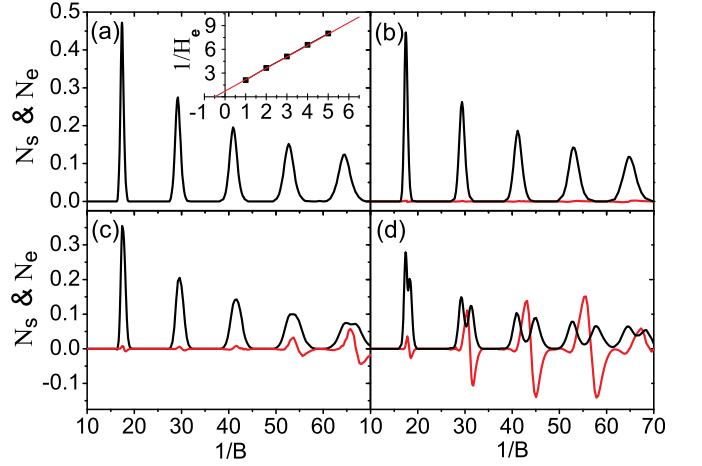}\\
\caption{(color online) $N_e$ (black) and $N_s$ (red or gray) vs.
${1}/{B}$ for different $V_R=0$ (a), $0.02$ (b), $0.05$ (c), and
$0.1$ (d). The other parameters are $E_F=-3$, the size of the center
region $L=40a$, and $\mathcal{T}=0.01$. In the inset of (a), the
dots are the inverse of peaks' maxima vs. $N\mathrm{th}$ peak, and
the line is $(N+\frac{1}{2})/\mathrm{ln}2$ vs. $N$.}
\end{figure}

Considering a small temperature gradient $\Delta \mathcal{T}$ and
zero bias applied on the longitudinal lead-1,3, we can set the
temperatures $\mathcal{T}_1 =\mathcal{T} +\Delta \mathcal{T}/2$,
$\mathcal{T}_3 =\mathcal{T} -\Delta \mathcal{T}/2$, $\mathcal{T}_2=
\mathcal{T}_4 =\mathcal{T}$, and the biases $V_1=V_3=0$. From the
open boundary condition with $J_{2e}=J_{4e}=0$, the transverse
voltage $V_2$ and $V_4$ can be obtained, and consequently the Nernst
coefficient $N_e=({V_2-V_4})/{\Delta \mathcal{T}}$. In the clean
system, $N_e$ is expressed as
\begin{eqnarray}
{N}_e=\frac{1}{e \mathcal{T}}\frac{\int dE ~
({T}_{21}-{T}_{23})(E-E_F) f(1-f)}{\int dE
~({T}_{21}+{T}_{23}+2{T}_{24} )f(1-f) },\label{Ne}
\end{eqnarray}
where $T_{2p} = T_{2\uparrow,p} + T_{2\downarrow,p}$. The spin
Hall current $J_{2s}$ and $J_{4s}$ are calculated with the closed
boundary condition having $V_2=V_4=0$. In the clean system,
$J_{2s}=-J_{4s}$ because of the symmetry property of the system.
In a dirty system, $J_{2s}$ may not equal to $-J_{4s}$ for a given
disorder configuration, but $J_{2s} =-J_{4s}$ still holds after
average over many configurations. The spin Nernst coefficient $N_s
\equiv J_{2s}/{\Delta \mathcal{T}}$, and can be reduced to:
\begin{eqnarray}
N_s=\frac{1}{4\pi}\int{dE} (\Delta T_{23}-\Delta T_{21})
 \frac{E-E_F}{k_B\mathcal{T}^2}f(1-f), \label{Ns}
\end{eqnarray}
where $\Delta T_{2p} = T_{2\uparrow,p}- T_{2\downarrow,p}$. At low
temperature limit ($\mathcal{T} \rightarrow 0$), the Nernst
coefficient $N_e$ and the spin Nernst coefficient $N_s$ usually
depend linearly on temperature. But while $\frac{\partial
E}{\partial
 ({T}_{21}-{T}_{23})}|_{E=E_F} =0$ ($\frac{\partial
E }{\partial (\Delta {T}_{21}-\Delta {T}_{23} )}|_{E=E_F}=0$), or in
other words ${T}_{21}-{T}_{23}$ $(\Delta {T}_{21}-\Delta {T}_{23})$
at $E=E_F$ is discontinuous, $N_e$ ($N_s$) is temperature
independent.

\begin{figure}
\includegraphics[bb=111 3 468 372, scale=0.50, clip=]{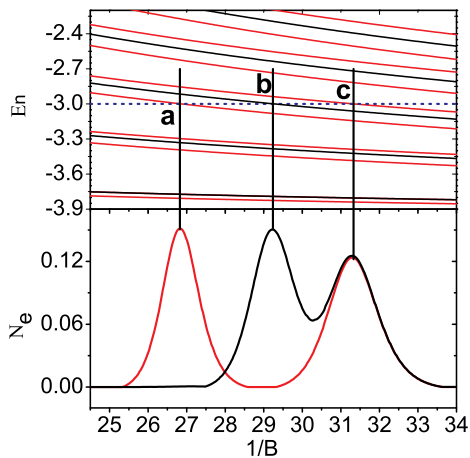}\\
\caption{(color online) (up panel) The LLs $E_n$ vs. $1/B$ for
$V_R=0$ (black) and $0.1$ (red or gray). The blue dotted line is the
Fermi level $E_F$. (down panel) $N_e$ vs. ${1}/{B}$ for the Rashba
SOI in lead-$2,4$ is 0 (black) and $0.1$ (red or gray). The
parameters are $E_F=1.0$, $V_R=0.1$, $L=20a$, and $
\mathcal{T}=0.01$. }
\end{figure}

In the numerical calculations, we set $t=\hbar^2/(2m^*a^2)$ as the
energy unit and $\frac{e}{h}a^2$ as the unit of the magnetic field
$B$. If taking the effective electron mass $m^*=0.05m_e$ and the
lattice constant $a=12.5nm$, $t$ is about $5meV$, $B=1$
corresponds to $4.2$ Tesla, and $V_R=0.1t$ corresponds to the
Rashba SOI parameter $\alpha=1.25\times10^{-11}\mathrm{eV}\cdot
\mathrm{m}$ which can be experimentally modulated by the gate
voltage. We consider square samples and the center-region size is
either $L=40a$ or $L=20a$ in our calculations. Temperature is
fixed at $\mathcal{T}=0.01$, that is about $1K$.

Fig.2 shows the Nernst coefficient $N_e$ and spin Nernst coefficient
$N_s$ versus the inverse of magnetic field $1/B$ for the different
SOI strength $V_R$ in the clean system ($W=0$). While without the
SOI ($V_R=0$), $N_s$ is exactly zero, but $N_e$ exhibits a series of
equal spacing peaks. $N_e$ peaks when the Fermi level $E_F$ crosses
over LLs, and it is damped when $E_F$ lies between adjacent LLs. The
peak interval is $e\hbar/(m^*E_F^*)$ where $E_F^*=E_F+4t$ is the
distance from the Fermi energy to the band bottom $-4t$. The inverse
of the height $H_e$ of the $N$th peak is linearly dependent on $N$,
with $H_e \propto N+1/2$ (as shown in inset of Fig.2a). Let us
explain these characteristics with aids of the physical picture in
Fig.1b. Under a strong $B$, the transmission coefficients
$T_{23}(E)$ and $T_{24}(E)$ are usually zero, and $T_{21}(E)$ is an
integer. Then the current $J_{2e}=(e/h)\int dE
T_{21}(E)[f_{2}(E)-f_{1}(E)]$ from Eq.(2). Due to the thermal
gradient, $f_{2}-f_{1}$ exhibits an oscillatory structure around
$E_F$ as shown in Fig.1b, and the electrons with energy above and
below $E_F$ contribute opposite signs to the thermocurrent $J_{2e}$.
When all LLs are far from $E_F$, $T_{21}$ is a constant near $E_F$,
the currents flowing in or out cancel each other, leading to
$J_{2e}=0$ at $V_2=0$. On the other hand, when $N$ LLs are below
$E_F$ but one LL is at $E_F$ (see Fig.1b), a net current $J_{2e}$ is
induced at $V_2=0$. In the open circuit case, $V_2$ has to be raised
to make $J_{2e}=0$, and $V_2$ is the ratio $1/(2N+1)$. In fact, from
Eq.(2) and assuming that $T_{32}=T_{42}=0$ and $T_{21}$ is an
integer for large $B$, we can analytically obtain that the peak
height of the Nernst coefficient is $H_e =
\frac{k_B}{e}{ln2}/(N+1/2)$. This result is identical with the
result of the thermopower in a two-terminal system.\cite{ref16}

\begin{figure}
\includegraphics[bb=80 13 455 443, scale=0.50, clip=]{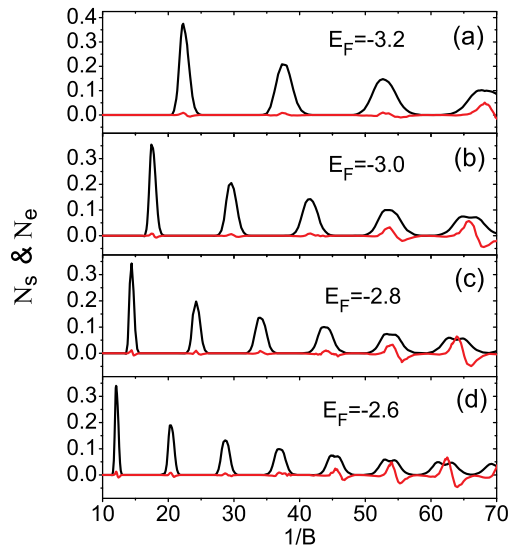}\\
\caption{(color online) $N_e$ (black) and $N_s$ (red or gray) vs.
$1/B$ for different Fermi level $E_F$ with the parameters
$V_R=0.05$, $\mathcal{T}=0.01$, and $L=40a$.}
\end{figure}

While in the presence of a SOI ($V_R\not=0$), the LLs split. As a
result, the peaks of the Nernst coefficient $N_e$ also split and the
spin Nernst coefficient $N_s$ emerges (see Fig.2). The splitting is
more pronounced for stronger SOI $V_R$ or weaker magnetic field $B$.
The positions of the right sub-peaks of $N_e$ are consistent with
LLs, but not the left sub-peaks. To see this, we magnify the second
peak of Fig.2d, and also plot in Fig.3 the LLs versus $1/B$ without
SOI ($V_R=0$) and with SOI ($V_R\not=0$). It clearly shows that the
left sub-peak is in line with the original un-split LL at $V_R=0$
(see mark b in Fig.3), not in alignment with the split LLs. In order
to thoroughly study the peak positions, we also plot $N_e$ for the
uniform system, in which SOI exists in all parts, including the
leads-$2,4$. Now the left sub-peak moves to align with the split LL.
(see mark a in Fig.3). So the counterintuitive phenomena entirely
comes from the non-uniformity of SOI, in which the SOI is absent in
the leads-$2,4$ and an interface between $V_R=0$ and $V_R\neq0$
emerges. This interface causes additional scattering for an incident
electron, and one of the edge states goes directly from lead-$1$ to
lead-$3$ instead of from lead-$1$ to lead-$2$, so the left sub-peak
position in $N_e$ is moved. This means that the Nernst effect can
reflect the detailed structure of the transverse leads and its
contact to the sample. This is essential difference to the regular
Hall effect.\cite{ref7,ref8,ref9}

\begin{figure}
\includegraphics[bb=10 70 582 362, scale=0.450, clip=]{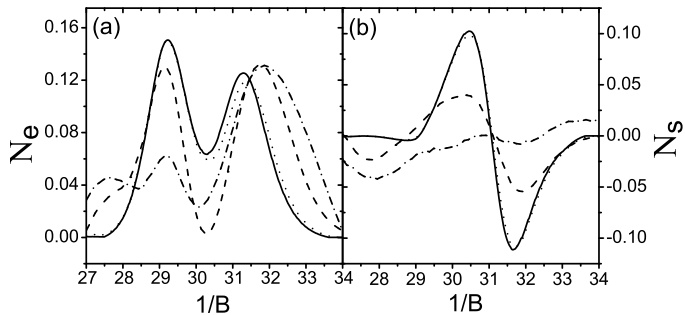}\\
\caption{$N_e$ (a) and $N_s$ (b) vs. $1/B$ for the different
disorder strengths $W=0$ (solid curve), $0.1$ (dotted curve), $0.5$
(dashed curve), and $1.0$ (dash-dotted curve). The other parameters
are $E_F=-3$, $V_R=0.1$, $\mathcal{T}=0.01$, and $L=20a$. }
\end{figure}

Next, we study the spin Nernst coefficient $N_s$, which emerges with
$V_R\not=0$ (see Fig.2b-d). In the vicinity of the right sub-peak of
$N_e$, $N_s$ exhibits an oscillatory structure and $N_s \propto
\partial [ln N_e(E_F)] /\partial E_F$. This relation of $N_s$ and
$N_e$ is similar to the semiclassical Mott relation between the
thermopower and conductance.\cite{ref16} However, $N_s$ is quite
small and does not show an oscillatory structure around the left
sub-peak of $N_e$, and the Mott-like relation breaks there. We can
qualitatively analyze these phenomena using Schr\"{o}dinger
equation, in which we can analytically obtain the split LLs and
corresponding wave functions. It is found that for weak SOI (such
as $V_R<0.1t$), the wave functions of the high sub-LLs are
strongly spin polarized in the z-direction, while one of the low
sub-LLs are hardly spin polarized. As a result, the spin current
$J_{2,4s}$ with an oscillatory structure only exists when $E_F$
crosses over the high sub-LLs, which corresponds to the positions
of the right sub-peaks of $N_e$. In addition, a larger SOI or a
weaker magnetic field will cause a stronger spin Nernst signals
(see Fig.2), due to the competition between the magnetic field and
the SOI. In fact, the Rashba SOI is to drive the electrons with
opposite spins to opposite directions transversely which leads to
the spin Nernst effect, but a magnetic field $B$ is to drive all
electrons in the same transverse direction. Thus, $B$ weakens the
spin Nernst effect.

The relation of the LLs $E_N$ with the magnetic field $B$ is: $E_N
= \frac{eB\hbar}{m^*}(N+1/2) -4t$, so the period in the Nernst
signal is ${e\hbar}/(m^*E_F^*)$, changeable by adjusting the Fermi
level $E_F$. In Fig.4, we plot $N_e$ and $N_s$ as functions of the
inverse of $B$ for different $E_F$. The results for different
$E_F$ show similar behaviors. With increasing of $E_F$, the peaks
of $N_e$ and the oscillatory structures of $N_s$ are getting
closer, and the magnitudes of $N_e$ and $N_s$ signals at same
filling factors are getting weaker.

Finally we discuss the disorder effect on the Nernst and spin
Nernst effects. Fig.5 shows $N_e$ and $N_s$ versus the inverse $B$
for different disorder strength $W$. Here $N_e$ and $N_s$ are
averaged over 500 disorder configurations. For a small disorder
(e.g. $W=0.1$), both $N_e$ and $N_s$ are hardly affected. For an
intermediate disorder, such as $W=0.5$, the two sub-peak heights
of $N_e$ almost keep their strengths as for $W=0$, but the valley
between two sub-peaks is greatly deepened, so that the two
sub-peak structure is even clearer (see Fig.5a). With further
increasing of the disorder $W$, the left sub-peak of $N_e$ is
decreased while the right sub-peak is less affected, meanwhile the
oscillatory structure of $N_s$ is weakened. However, in the
vicinity of the left sub-peak of $N_e$, the spin Nernst
coefficient $N_s$, which is very small at $W=0$, is enhanced by
$W$ (see Fig.5b). Finally, for very large disorder $W$, the system
goes into an insulating regime, both $N_e$ and $N_s$ vanish.

Before summary, we would like to make a couple comments concerning
the novel spin Nernst effect. (i) The spin Nernst effect is NOT a
simple combination of the Seebeck effect and the spin Hall effect.
In fact, the Seebeck coefficient is mainly determined by
$(dT_{13}(E))/(dE)|_{E=E_F}$, while the spin Nernst coefficient
depends on $\Delta T_{23}$ and $\Delta T_{21}$. (ii) The spin Nernst
effect can be measured in similar ways that the spin Hall effect is
observed,\cite{ref5,ref6} e.g. through spin accumulations.

In summary, the Nernst effect and spin Nernst effect in a
two-dimensional cross-bar with a spin-orbit interaction and under
a perpendicular magnetic field are investigated. The Nernst signal
exhibits a series of peaks, and the inverse of a peak height goes
linearly to the sequence number of the peak. While in the presence
of a SOI, these Nernst peaks split, and the spin Nernst effect
appears, which exhibits an oscillatory structure versus the
magnetic field. The relation of the Nernst and spin Nernst
coefficients is similar to the semiclassical Mott relation around
one sub-peak, but has a great discrepancy around the other
sub-peak. In addition, the disorder effect on the Nernst and spin
Nernst effects is also discussed.

{\bf Acknowledgments:} We gratefully acknowledge the financial
support from the Chinese Academy of Sciences, US-DOE under Grant No.
DE-FG02-04ER46124 and US-NSF, and NSF-China under Grant Nos.
10525418, 60776060, and 10734110.

\end{document}